# Momentum-dependent charge-density-wave gap formation in ZrTe$_{2.98}$Se$_{0.02}$

Iori Ishiguro[1], Hayate Kunitsu[1], Natsuki Mitsuishi[1], Shunsuke Tsuda[2], Koichiro Yaji[2,3], Yoichi Yamakawa[1], Hiroshi Kontani[1], Takahiro Shimojima[1]

[1] *Department of Physics, Nagoya University, Furo-cho, Nagoya 464-8602, Japan*

[2] *Center for Basic Research on Materials, National Institute for Materials Science, Tsukuba 305-0047, Japan*

[3] *Unprecedented-scale Data Analytics Center, Tohoku University, Sendai 980-8578, Japan*

**We investigated the energy gap formation across the charge density wave (CDW) transition in ZrTe$_{2.98}$Se$_{0.02}$. By employing laser photoemission microscopy, we clearly resolved one elliptical Fermi surface (FS) around the Brillouin zone (BZ) center, and two quasi-one-dimensional FSs along the BZ boundary. We further mapped the intensity difference between the FSs below and above the CDW transition temperature. We found that the CDW gap formation is limited to the momentum region 0.25 Å$^{-1}$ < $k_y$ < 0.8 Å$^{-1}$ along $\overline{\text{B}}$-$\overline{\text{D}}$ line, which coincides with the location of one of the quasi-one-dimensional FSs. Characteristic momentum dependence in the CDW gap suggests the importance of both FS nesting and band-dependent electron-phonon coupling for understanding the CDW state in ZrTe$_3$ system.**

The interplay between charge density waves (CDW) and superconductivity in low-dimensional materials has been a major issue in condensed matter physics.[1] The CDW transition is generally explained by the Peierls mechanism that the nesting of the Fermi surface (FS) leads to an electronic instability and induces a periodic lattice distortion (PLD) through the electron-phonon coupling (EPC).[2] However, recent studies indicate that the CDW transition cannot be fully explained by the simple Peierls scenario.[3,4] Instead, the momentum ($k$) -dependent EPC has been also considered as the origin of the CDW.[5–7]

ZrTe$_3$ is a quasi-one-dimensional (q1D) material that exhibits filamentary superconductivity below $T_c \sim 2$ K at ambient pressure.[8,9] The CDW transition temperature ($T_{CDW}$) of 63 K[8,10] is characterized by an incommensurate PLD modulation of $q_{CDW} \sim (0.07, 0, 0.333)$.[11] The crystal structure is monoclinic with trigonal prismatic ZrTe$_6$ chains running along the $b$-axis, belonging to the space group $P2_1/m$ [Fig. 1(a)].[11] Se doping suppresses the CDW transition and induces bulk superconductivity at 4 K.[12]

Figure 1(b) shows the three-dimensional Brillouin zone (BZ) and the two-dimensional BZ projected onto the (001) surface of ZrTe$_3$ which corresponds to the easy cleavage plane. The first-principles band calculations predicted that the elliptical FS consists of a hole band centered at $\bar{\Gamma}$ point and the q1D FSs are formed by the electron bands extending along $\bar{B}$-$\bar{D}$ line.[13,14] The overall view of the electronic structure was supported by the previous angle-resolved photoemission spectroscopy (ARPES).[15–20] Especially, the synchrotron ARPES measurements suggested that the q1D FSs are separated in the middle of $\bar{B}$-$\bar{D}$ points while they are nearly degenerated around $\bar{D}$ point.[17] Temperature ($T$) -dependent ARPES study reported that the spectral weight at $E_F$ around $\bar{D}$ point is reduced well above $T_{CDW}$,[15] suggesting a pseudogap behavior in the

q1D FSs. On the other hand, the CDW gap was not observed at $\bar{\text{B}}$ point[15] and at a certain $k$ intermediate between $\bar{\text{B}}$ and $\bar{\text{D}}$ points.[18] The $k$ dependence of the CDW gap formation in the q1D FSs has been an important issue for understanding the origin of the CDW in ZrTe$_3$ system. To resolve this issue, it is crucial to continuously map out the CDW gap formation in $k$ space.

In this letter, we determined the shape of the FSs in ZrTe$_{2.98}$Se$_{0.02}$ by using laser photoemission microscopy. We further examined the intensity difference between the FSs below and above $T_{\text{CDW}}$. We found the CDW gap formation at 0.25 Å$^{-1}$ < $k_y$ < 0.8 Å$^{-1}$ along $\bar{\text{B}}$-$\bar{\text{D}}$ line, which coincides with the location of one of the q1D FSs. The peculiar $k$-dependence of the CDW gap in ZrTe$_{2.98}$Se$_{0.02}$ can be explained by the FS nesting between q1D FSs and band-dependent electron-phonon coupling.[21,22]

Single crystals of ZrTe$_{3-x}$Se$_x$ were purchased from HQ Graphene. The Se composition $x$ was determined to be ~0.02 by energy-dispersive X-ray spectroscopy, which shows $T_{\text{CDW}}$ of ~30 K.[12] The FS observations were performed at the National Institute for Materials Science (NIMS) using an imaging-type spin-resolved photoemission microscopy (iSPEM)[23-25] equipped with a 10.9 eV circularly polarized laser. The combination of high spatial resolution (a detecting area of 10 × 30 μm$^2$) and a small $k_z$ broadening owing to the bulk sensitivity of 10.9 eV laser enables a detailed investigation of the FS with high data quality. The laser photoemission microscopy measurements were conducted under 1.2 × 10$^{-8}$ Pa and the samples were cleaved *in situ* at room temperature. Density functional theory (DFT) calculations were performed using the crystal structure parameters reported in Ref.12.

First, we show the FS mapping at 20 K in Fig. 1(c) obtained by summing the data acquired by left- and right-circularly polarized (LCP and RCP) lasers. In addition to the elliptical FS "α" at $\bar{\Gamma}$ point, the q1D FSs were observed with nonmonotonic distribution of the FS intensity along $\bar{B}$-$\bar{D}$ line. These observations are consistent with previous ARPES studies.[15–20] In order to analyze the shape of the FSs, we performed momentum distribution curve (MDC) analysis. Figure 2(a) shows the peak positions of the MDCs for the FS α around $\bar{\Gamma}$ point. The red (blue) markers represent the peak positions of the MDC along $k_y$ ($k_x$). Examples of the analysis are exhibited in Fig. 2(b) which indicates the shift of the Fermi momentum ($k_F$) forming the elliptical FS. Although previous ARPES studies reported a splitting of the FS α,[17,19] it was not observed in this study.

Next, we track the $k_F$ of the q1D FSs along $\bar{B}$-$\bar{D}$ line [Fig. 2(c)]. One can see the two-peak structure in the MDCs along $k_x$ (blue markers) as demonstrated at the representative $k_y$ of –0.30 Å$^{-1}$, –0.02 Å$^{-1}$, 0.40 Å$^{-1}$ and 0.78 Å$^{-1}$ in Fig. 2(d), indicating the q1D nature of the FSs around the BZ boundary. Distance between the two peaks (i.e. 2$k_F$ of the q1D FSs) are estimated to be 0.07 Å$^{-1}$ - 0.11 Å$^{-1}$. Here we focus on the MDCs along $k_y$ in Fig. 2(e). It is notable that the MDCs along $k_y$ show the peak structures suggesting that the q1D FSs have a curvature especially around $k_y$ ~ ±0.25 Å$^{-1}$. As summarized in Fig. 2(f), the red markers suggest that the one of the q1D FSs closes at $k_y$ ~ ±0.25 Å$^{-1}$. On the other hand, the peak structure around |$k_y$| < 0.1 Å$^{-1}$ might indicate the warping of the q1D FS around $\bar{B}$ point. These observations are consistently understood by the schematic FSs in Fig. 2(g). The inner q1D FS "β" (thin black curves) exists around $\bar{D}$ point and closes at $k_y$ ~ ±0.25 Å$^{-1}$, while the outer q1D FS "γ" runs along $k_y$ with slightly rounded shape around $\bar{B}$ point.

We compare the FSs with those calculated by the DFT with spin-orbit coupling in Fig. 3 (a-d). First, we found that the number of the FS sheets is well reproduced by the calculations. The orbital components were indicated by the size of the markers. The FS α has the main contributions from Zr 4*d* and Te 5*p* [Te(1) site] orbital components, while the q1D FSs β and γ mainly consist of Te 5*p* [Te(2) and Te(3) sites] orbitals. Second, we notice that the shape of the calculated FS α is slightly asymmetric, i.e. the right side of the FS is more pointed, at $k_z$ = 0.625 π and 0.725 π, which is also found in the experimental FS α [Figs. 2(f) and 2(g)]. Third, the DFT calculations predict the large $k_z$ dependence around $\bar{B}$ point for the FS γ (outer q1D FS). The outer q1D FS is continuous around $\bar{B}$ point at $k_z$ = 0.525 π and 0.625 π, being consistent with the observation [Figs. 2(f) and 2(g)]. These comparisons reveal that the experimental FSs are the most reproduced by the calculated ones at $k_z$ = 0.625 π [Fig. 3(b)]. These results suggest that the present observations capture the intrinsic electronic properties of ZrTe$_{2.98}$Se$_{0.02}$.

Finally, we mapped the intensity difference between the FSs below (20 K) and above (70 K) $T_{CDW}$ in Fig. 4(a) where the blue (red) signal indicates the higher intensity at 20 K (70 K). A striking feature in Fig. 4(a) is the elliptical red signal surrounded by the blue signal around $\bar{\Gamma}$ point. This result is understood by considering both the quasiparticle peak enhancement and enlargement of the FS volume toward low *T* as reported by the previous ARPES.[15, 19] Another important feature can be found in the q1D FSs. The FS intensity is reduced around $\bar{D}$ point (red) at 20 K while it is enhanced near $\bar{B}$ point (blue). We note that the shape and location of the q1D FSs remain constant with *T* change.[17] In this case, the intensity-difference analysis can extract the *k*-dependence of the CDW gap signature. These observations for the q1D FSs are, therefore, consistent with the previous ARPES reporting the energy gap at $\bar{D}$ point and the quasiparticle peak enhancement at $E_F$ (no gap) at $\bar{B}$ point toward low *T*.[15] We further found that the CDW gap formation in ZrTe$_{2.98}$Se$_{0.02}$ is limited to the *k* region of ~0.25 Å$^{-1}$ < $k_y$ < 0.8 Å$^{-1}$ along $\overline{\text{B}}$-$\overline{\text{D}}$ line. In Figs. 4(b) and 4(c), the schematic FSs [Fig. 2(g)] are superimposed on the enlarged images along the $\overline{\text{B}}$-$\overline{\text{D}}$ line in Fig. 4(a). It is remarkable that the $k$ region showing the intensity suppression at low $T$ (red) coincides with the location of the FS β (thin black curve).

It is widely believed that the lattice periodicity in the CDW state has a close relation to the FS nesting condition. The $2k_F$ value along $k_x$ for the q1D FSs in this study is indeed consistent with the PLD modulation $\boldsymbol{q}_{CDW}$ ~ (0.07, 0, 0.333) reported by the electron diffraction measurements.[11] On the other hand, the CDW gap formation in Fig. 4(a) shows the large $k_y$ dependence, i.e. the sign of the intensity difference switches at $k_y$ ~ ±0.25 Å$^{-1}$. This observation seems not to be explained solely by the above FS nesting. A possible explanation is the orbital-selective CDW gap formation as reported for TaTe$_4$ by ARPES.[26] In this study, the CDW gap mainly opens in the electronic bands of the out-of-plane orbitals, while the in-plane orbitals contribute to the metallic state. Our DFT calculations in Fig. 3(b) show that the orbital components for the FSs β and γ gradually change along $k_y$ direction. The Te $5p$ orbital component from the Te(2) and Te(3) sites which contribute to the one-dimensional chain in the CDW state, is dominant around $\overline{\text{D}}$ point. On the other hand, the Te $5p$ orbital of Te(1) site and Zr $4d$ orbital component which have three-dimensional nature, increase toward $\overline{\text{B}}$ point. These gradual change in the orbital components seem to be consistent with the overall trend of the CDW gap distribution in Fig. 4(a).

Important insight from the present results is the correspondence between the locations of the FS β and the CDW gap region. A plausible scenario is the band-dependent electron-phonon coupling.[21,22] Ref.21 reported the calculations of the electron-phonon coupling matrix element for each FS sheet. Two phonon modes, being sensitive to the CDW gap formation probed by the Raman scattering measurements, couple to the electrons around $\bar{\text{D}}$ point. The location of the FS β nearly coincides with the $k$ region where the electron-phonon coupling is prominent. The present results suggest that the FS nesting condition and the band-dependent electron-phonon coupling can be both necessary for understanding the CDW state in ZrTe$_3$ system.

In conclusion, we investigated the energy gap formation across the CDW transition of ZrTe$_{2.98}$Se$_{0.02}$ by employing laser photoemission microscopy. We observed the FS α around $\bar{\Gamma}$ point, and the FSs β and γ along $\bar{\text{B}}$-$\bar{\text{D}}$ line. We successfully detected the intensity difference between the FSs below and above $T_{\text{CDW}}$. The CDW gap formation was limited to the $k$ region of ~0.25 Å$^{-1}$ < $k_y$ < 0.8 Å$^{-1}$ along $\bar{\text{B}}$-$\bar{\text{D}}$ line, which coincides with the location of the FS β. The peculiar $k$-dependence in the CDW gap formation of ZrTe$_{2.98}$Se$_{0.02}$ is better explained by considering both FS nesting and band-dependent electron-phonon coupling.

**Acknowledgement**

The authors thank F. Arai for the technical support of iSPEM measurements. The present work was partially supported by JST CREST, Japan (Grant No JPMJCR2435), the Japan Society for the Promotion of Science KAKENHI (Grant Nos. 24K01352, 24K17591, 25K07195), and the Innovative Science and Technology Initiative for Security Grant Number JPJ004596, ATLA, Japan. This work was supported by "Advanced Research Infrastructure for Materials and Nanotechnology in Japan (ARIM)" of the Ministry of Education, Culture, Sports, Science and Technology (MEXT) (Proposal Number JPMXP1225NU0408).

**References:**

1. P. Monceau, *Advances in Physics* **61**, 325–581 (2012).

2. J.-P. Pouget, *Comptes Rendus Physique* **17**, 332–356 (2016).

3. F. Schmitt, P. S. Kirchmann, U. Bovensiepen, R. G. Moore, J.-H. Chu, D. H. Lu, L. Rettig, M. Wolf, I. R. Fisher, and Z.-X. Shen, *New J. Phys.* **13**, 063022 (2011).

4. X. Zhu, J. Guo, J. Zhang, and E. W. Plummer, *Advances in Physics: X* **2**, 622–640 (2017).

5. M. D. Johannes, I. I. Mazin, and C. A. Howells, *Phys. Rev. B* **73**, 205102 (2006).

6. M. D. Johannes, and I. I. Mazin, *Phys. Rev. B* **77**, 165135 (2008).

7. K. Rossnagel, *J. Phys.: Condens. Matter* **23**, 213001 (2011).

8. S. Takahashi, T. Sambongi, and S. Okada, *J. Phys. Colloques* **44**, C3-1736 (1983).

9. S. Tsuchiya, K. Matsubayashi, K. Yamaya, S. Takayanagi, S. Tanda, and Y. Uwatoko, *New J. Phys.* **19**, 063004 (2017).

10. S. Takahashi, T. Sambongi, J. W. Brill, and W. Roark, *Solid State Communications* **49**, 1031–1034 (1984).

11. D. J. Eaglesham, J. W. Steeds, and J. A. Wilson, *J. Phys. C: Solid State Phys.* **17**, L697 (1984).

12. X. Zhu, W. Ning, L. Li, L. Ling, R. Zhang, J. Zhang, K. Wang, Y. Liu, L. Pi, Y. Ma, H. Du, M. Tian, Y. Sun, C. Petrovic, and Y. Zhang, *Sci Rep* **6**, 26974 (2016).

13. K. Stöwe, and F. R. Wagner, *Journal of Solid State Chemistry* **138**, 160–168 (1998).

14. C. Felser, E. W. Finckh, H. Kleinke, F. Rocker, and W. Tremel, *J. Mater. Chem.* **8**, 1787–1798 (1998).

15. T. Yokoya, T. Kiss, A. Chainani, S. Shin, and K. Yamaya, *Phys. Rev. B* **71**, 140504 (2005).

16. P. Starowicz, C. Battaglia, F. Clerc, L. Despont, A. Prodan, H. J. P. van Midden, U. Szerer, A. Szytuła, M. G. Garnier, and P. Aebi, *Journal of Alloys and Compounds* **442**, 268–271 (2007).

17. M. Hoesch, X. Cui, K. Shimada, C. Battaglia, S. Fujimori, and H. Berger, *Phys. Rev. B* **80**, 075423 (2009).

18. M. Hoesch, L. Gannon, K. Shimada, B. J. Parrett, M. D. Watson, T. K. Kim, X. Zhu, and C. Petrovic, *Phys. Rev. Lett.* **122**, 017601 (2019).

19. S.-P. Lyu, L. Yu, J.-W. Huang, C.-T. Lin, Q. Gao, J. Liu, G.-D. Liu, L. Zhao, J. Yuan, C.-T. Chen, Z.-Y. Xu, and X.-J. Zhou, *Chinese Phys. B* **27**, 087503 (2018).

20. M. Maki, Y. Fujimoto, I. Yamamoto, J. Azuma, S. Kobayashi, S. Demura, and H. Sakata, *J. Phys. Soc. Jpn.* **91**, 094708 (2022).

21. Y. Hu, F. Zhang, X. Ren, J. Feng, and Y. Li, *Phys. Rev. B* **91**, 144502 (2015).

22. J. Diego and M. Calandra, Preprint at 10.48550/arXiv.2602.04534 (2026).

23. K. Yaji and S. Tsuda, *e-J. Surf. Sci. Nanotechnol.* **22**, 46 (2024).

24. K. Yaji and S. Tsuda, *Sci. Technol. Adv. Mater. Methods* **4**, 2328206 (2024).

25. S. Tsuda and K. Yaji, *e-J. Surf. Sci. Nanotechnol.* **22**, 170 (2024).

26. R. Z. Xu, X. Du, J. S. Zhou, X. Gu, Q. Q. Zhang, Y. D. Li, W. X. Zhao, F. W. Zheng, M. Arita, K. Shimada, T. K. Kim, C. Cacho, Y. F. Guo, Z. K. Liu, Y. L. Chen, and L. X. Yang, *npj Quantum Mater*. **8**, 44 (2023).

Fig. 1

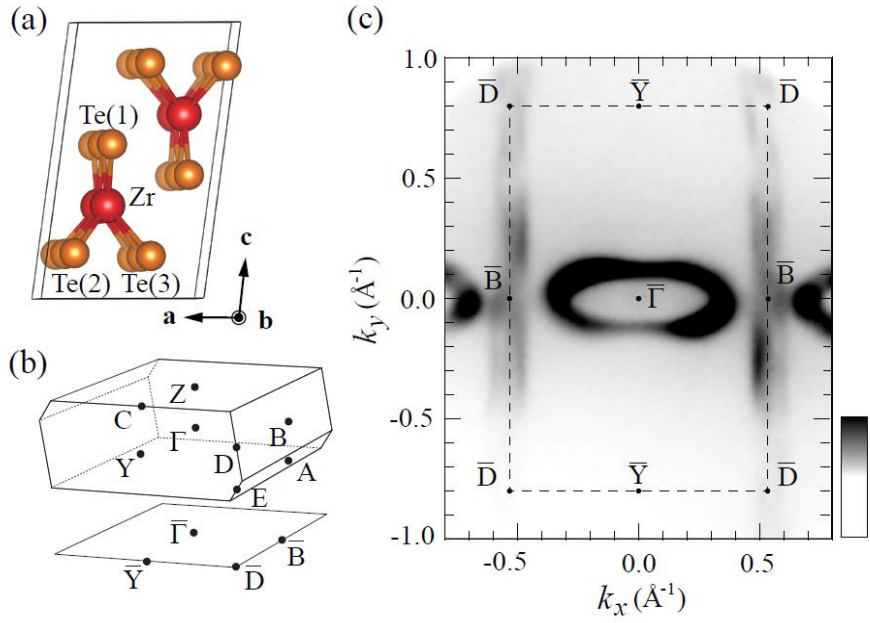

**Fig. 1.** Crystal structure and FSs of ZrTe$_3$. (a) Crystal structure of ZrTe$_3$. Red and orange spheres represent the Zr and Te atoms, respectively. (b) Three-dimensional BZ and the two-dimensional BZ projected onto (001) surface. (c) FS mapping of ZrTe$_{2.98}$Se$_{0.02}$ obtained at 20 K by summing the data measured by LCP and RCP lasers.

Fig. 2

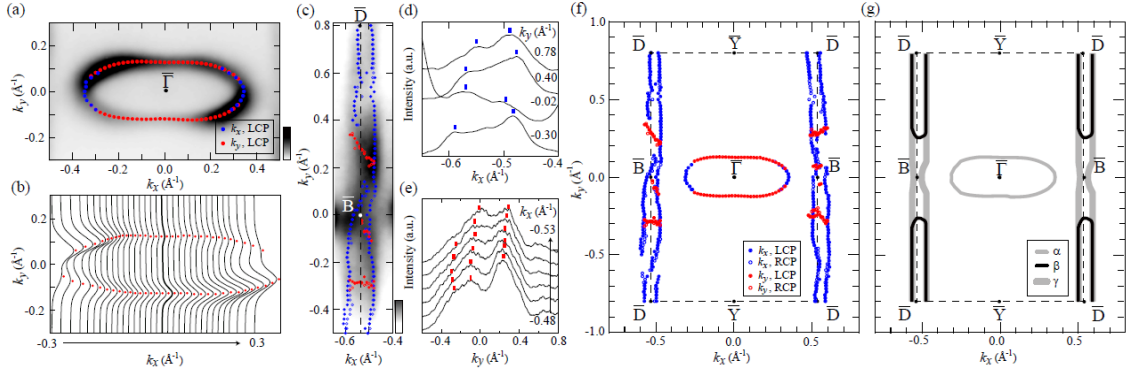

**Fig. 2**. Determination of the FS shape of ZrTe$_{2.98}$Se$_{0.02}$. (a) FS mapping around $\bar{\Gamma}$ point obtained by summing the data measured by LCP and RCP lasers at 20 K. Red (blue) circles indicate the $k_F$ obtained from the peak positions of MDCs along $k_y$ ($k_x$) direction. (b) MDCs along $k_y$ directions for the FS α. (c) FS mapping along $\bar{B}$-$\bar{D}$ line obtained by summing the data measured by LCP and RCP lasers. (d,e) MDCs of the data in (c) along $k_x$ and $k_y$ direction, respectively. (f) Summary of the $k_F$ obtained from the MDC peak positions. (g) Schematics of the experimental FSs.

Fig. 3

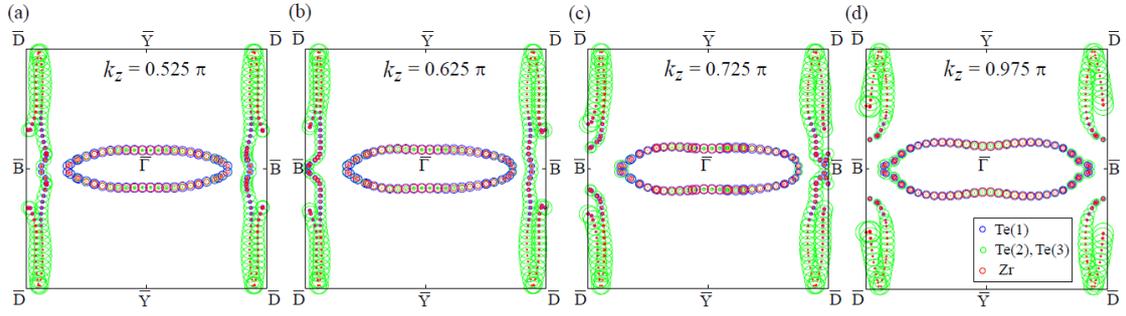

**Fig. 3**. (a-d) DFT calculations of the FS of ZrTe$_3$ at $k_z$ = 0.525π, 0.625π, 0.725π and 0.975π respectively. The size of the markers is proportional to the orbital contributions from Zr 4*d* and Te 5*p* orbitals. Note that the scale in *k* space has been adjusted so that the distances of $\overline{D}$-$\overline{B}$-$\overline{D}$ line and $\overline{D}$-$\overline{Y}$-$\overline{D}$ line are the same.

Fig. 4

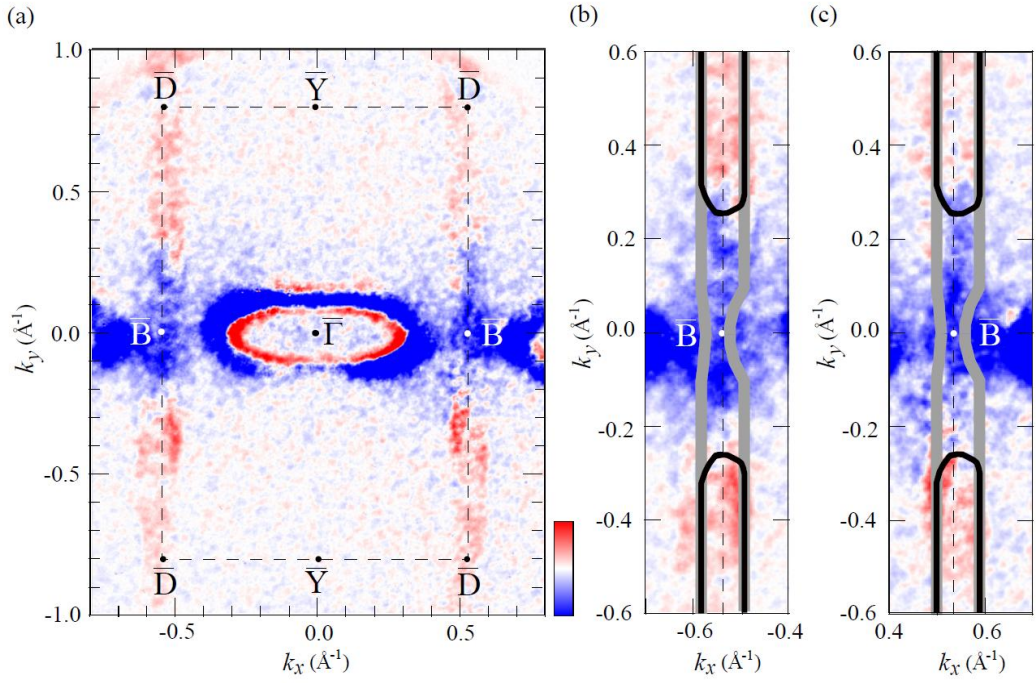

**Fig. 4.** (a) Intensity difference between the FSs of ZrTe$_{2.98}$Se$_{0.02}$ obtained at 70 K and 20 K. Red and blue colors represent higher intensities at 70 K and 20 K, respectively. (b,c) Enlarged images and schematic FSs along $\overline{B}$-$\overline{D}$ line for the left and right part of (a), respectively. The FS β (γ) is presented by thin black curves (thick gray curves).